# Incipient Metals: Functional Materials with a Unique Bonding Mechanism

Matthias Wuttig[1]*, Volker L. Deringer[2,3], Xavier Gonze[4], Christophe Bichara[5], Jean-Yves Raty[6]

**Affiliations:**

[1]Institute of Physics IA, RWTH Aachen University, 52074 Aachen, Germany

[2]Institute of Inorganic Chemistry, RWTH Aachen University, 52056 Aachen, Germany

[3]Engineering Laboratory, University of Cambridge, Cambridge CB2 1PZ, United Kingdom

[4]Institute of Condensed Matter and Nanosciences, Université Catholique de Louvain, 1348 Louvain-la-Neuve, Belgium

[5]Aix Marseille Univ, CNRS, CINAM, Campus de Luminy, Marseille, 13288, France

[6]CESAM & Physics of Solids, Interfaces and Nanostructures, B5, Université de Liège, B4000 Sart-Tilman, Belgium

*Correspondence to: wuttig@physik.rwth-aachen.de.

**Abstract**: While solid-state materials are commonly classified as covalent, ionic, or metallic, there are cases that defy these iconic bonding mechanisms. A prominent example is given by phase-change materials (PCMs) for data storage or photonics, which have recently been argued to show "resonant bonding"; a clear definition of this mechanism, however, has been lacking until the present day. Here we show that these solids are clearly different from resonant bonding in the π-orbital systems of benzene and graphene. Instead, they exhibit a unique mechanism between covalent and metallic bonding, which we call "metavalent" bonding. The materials are on the verge of electron delocalization, which explains their exceptional property portfolio, and we therefore argue that they represent "incipient metals". This yields deeper, fundamental insight into the bonding nature of solid-state materials, and is expected to accelerate the discovery and design of new functional materials including PCMs and thermoelectrics.

**One Sentence Summary:** A unique bonding mechanism is identified in a class of inorganic materials, located between covalent and metallic interactions, but clearly distinct from both.

**Main Text:** The structures and properties of materials are controlled by diverse interatomic interactions. Textbooks define the prototypical cases of covalent, ionic, and metallic bonding, but reality most often lies in between such idealized descriptions (*1*–*3*). Chemical bonding is not directly linked to quantum-mechanical observables, and has often been the topic of heated debates (*4*); no less, its understanding is a key requirement for the discovery and design of new materials. Over the recent years, predictive bonding models have enabled the rational design of magnetic intermetallics (*5*), supramolecular assemblies (*6*), or novel thermoelectrics (*7*). Today, quantum-mechanically and bond-theoretically driven high-throughput searches can point out target materials and compositions that had hitherto not been thought of (*8*–*10*).

Materials scientists have recently re-discovered strong interest in a mechanism called “resonant bonding” (RB) in inorganic solids (*11*, *12*), which has been linked to a portfolio of useful properties (*13*–*16*) but only described in empirical terms so far. The terminology goes back almost a century: Pauling’s early work in the 1930s (*17*) proposed “resonant” bonding in the benzene molecule using a valence-bond framework, a concept that is still part of every undergraduate organic-chemistry textbook. This idea was later transfered to solids and suggested as an inherent property of several IV–VI semiconductors, based on experimentally observable quantities to which we will return below (*11*, *12*). In parallel, a similar term is used for “resonating valence bond” (RVB) materials such as high-temperature superconductors (*18*). Clearly, the concept of resonance is invoked for very different material classes, and the question arises whether “resonant bonding” should be viewed as a distinct mode of bonding in solids at all.

In this work, we provide this missing definition, and thereby clarify the fundamental nature of “resonantly bonded” inorganic solids including chalcogenide phase-change materials (PCMs). We show that these materials depart considerably from covalent bonding in their behavior, approaching the metallic regime, yet have properties that differ significantly from both metals and covalently bonded compounds. We therefore suggest that they be referred to as “*metavalent*” solids or *incipient metals* instead.

We begin by asking how the bonding nature of solids can be described in empirical terms. Among the simplest bonding indicators is the distribution of electrons in a material. In textbook examples of covalent solids, say diamond or silicon, the electrons are localized near the nuclei and in the bonds between them. In ionic materials, electrons are localized as well, mostly near the anions, leading to electrostatic interactions and therefore to brittle and insulating behavior. By stark contrast, in metals, there are much fewer valence electrons per atomic neighbor (“electron deficiency”), and therefore those electrons are delocalized and readily move through the lattice. A direct consequence is the electronic conductivity, which is experimentally accessible and provides a first indication of a material’s bonding nature. In the following, we will assume ideally ordered, stoichiometrically pure, defect-free crystals, and experiments that come as close to this limit as possible. While defects and doping may also have a strong influence on the behavior of a material, this influence will be very specific for a given composition.

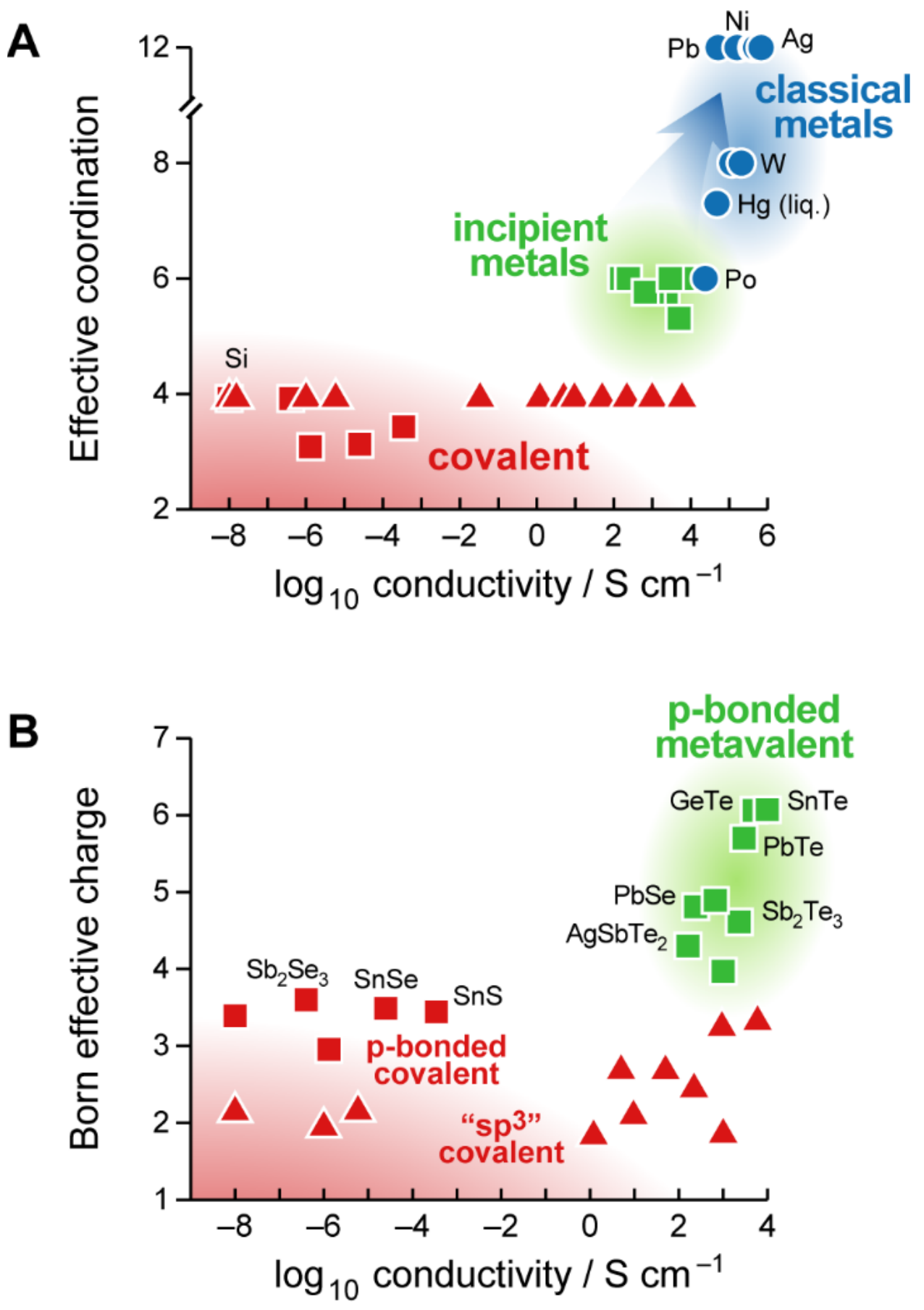

**Fig. 1**. Electronic and structural fingerprints of materials. (A) Effective coordination number in a range of solids from semiconductors to classical metals. “Resonantly bonded” solids such as GeTe are located between the regime of covalent semiconductors (red) and metals (blue), and we therefore suggest the term *incipient metals*. (B) Same for Born effective charges as a fingerprint of “resonant bonding”, which we here propose to call *metavalent* instead. In both viewgraphs, triangles denote “$sp^3$-bonded” (zincblende-like) crystals whereas squares denote “p-bonded” (orthorhombic or rocksalt-like) systems; note that the latter terminology refers to atomic structures only, and p-bonded systems are found to be either covalent or metavalent. Both quantities have been plotted versus the tabulated electric conductivity at or close to room temperature (Table S1).

A second defining feature of a material is its atomic coordination number, which is again linked to the bonding nature. In metals, where non-directional bonding prevails, atoms usually have eight (body-centered cubic) or twelve nearest neighbors (cubic or hexagonal close packing). In “classical” covalent solids, the coordination numbers are much lower, in accord with the 8–$N$ rule: silicon, an element from the fourth main group, forms four covalent bonds in its stable structure, whereas phosphorus, from the fifth main group, forms three. **Fig. 1A** now collects such data for a large set of elements and compounds, correlating effective coordination numbers (ECoN) (*19*) with conductivity. Metals are characterized by large coordination numbers and high conductivities (blue), while covalent semiconductors show low conductivities and ECoN = 4 for $sp^3$-bonded materials (such as Si or GaAs; red triangles in Figure 1a) or ECoN ≈ 3 for p-bonded systems (red squares). Materials that had previously been referred to as "resonantly" bonded are highlighted in green in this viewgraph. It becomes immediately apparent that they are intermediate between both realms, exceeding the coordination numbers prescribed by the 8–*N* rule, and approaching the characteristics of metals (blue). This fits well with the above-mentioned concept of electron deficiency: the number of valence electrons is the same for GeSe and GeTe, but the latter has a much higher ECoN, and therefore a lower valence electron count per atomic neighbor. In brief, we find materials such as GeTe to be what we call *incipient* ("beginning") metals.

The mere observation that materials become more metallic when moving down a group in the Periodic Table is not surprising at all. However, we find that this pathway is distinctly different in various materials classes. To this end, we inspect one of the characteristic fingerprints of "resonant bonding" (*11–13*): namely, the fact that they exhibit anomalously high Born effective charges (which characterize the sensitivity of a material to lattice distortions, that is, the chemical bond polarizability). We plot these for the same set of materials in **Fig. 1B**, again correlating the data with electric conductivity to trace the covalent → metallic transition. For $sp^3$-bonded systems (red triangles), including Si, GaAs, and other zincblende-type materials, the Born effective charge increases only slowly with increasing metallicity. Similarly, the coordination numbers remain constant (ECoN = 4 for Si, Ge, and Sn) before jumping rapidly (ECoN = 12 for Pb; Fig. 1A). By contrast, this transition looks quite different for many main-group chalcogenides and other p-bonded systems (red and green squares in Fig. 1): the coordination numbers increase gradually even in the presence of a band gap, incompatible with the 8–*N* rule, and the Born effective charges are atypically high (Fig. 1B).

How does this, now, relate to Pauling-like resonance in benzene—or its extended analogues, graphene and graphite? We will argue in the following that the bonding in these systems is very different from that in incipient metals. While benzene and graphite have a resonating system, they *also* contain strong covalent bonds, typically referred to as the "$sp^2$" system. The role of this strong "backbone" becomes obvious when bond indicators such as optical phonon frequencies and their pressure dependence are investigated.

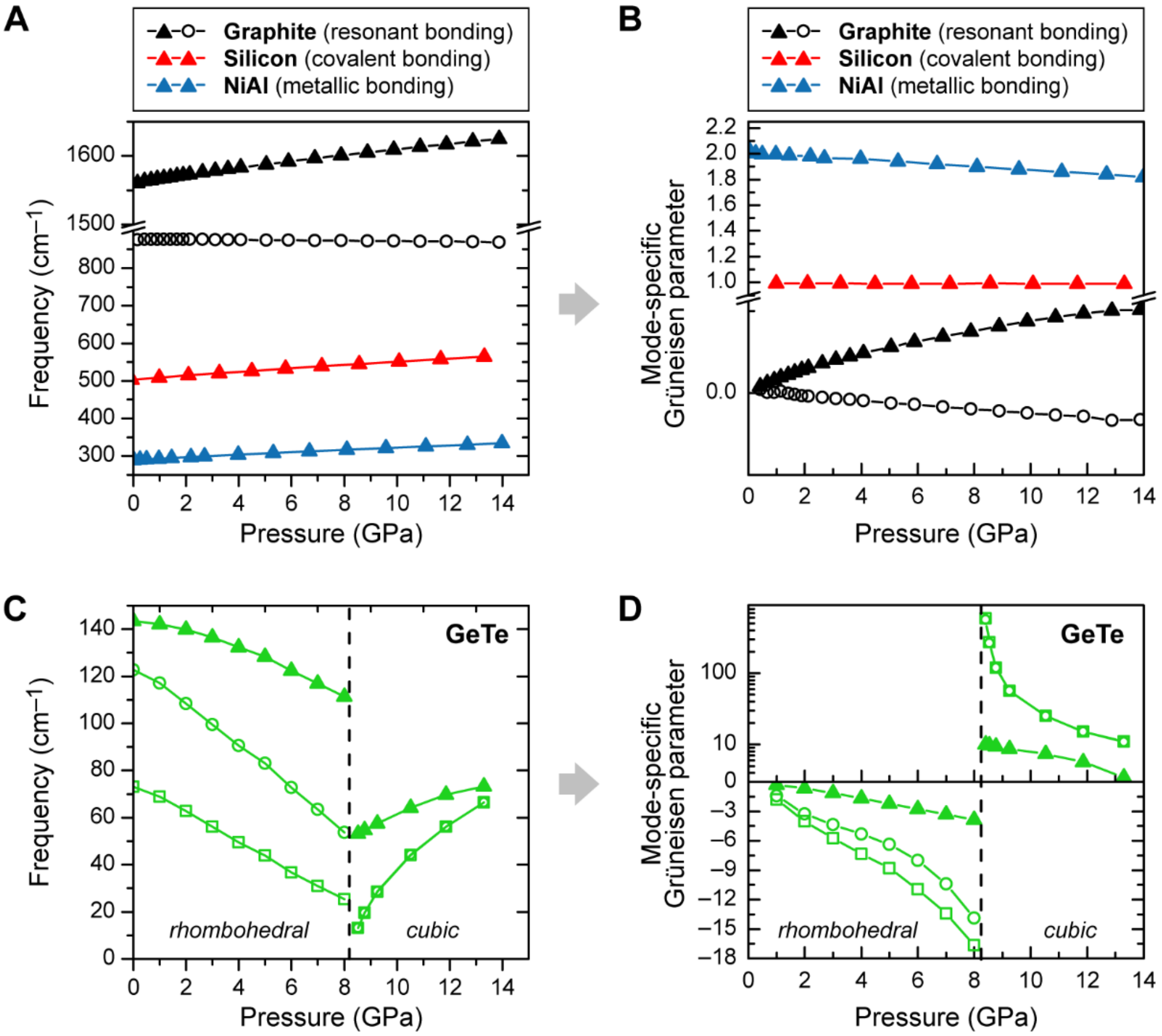

**Fig. 2**. Pressure response of the characteristic vibrational modes in materials, used here to gauge their bonding nature. This viewgraph compares data for graphite, for textbook examples of covalent (Si) and metallic materials (NiAl), and for the incipient metal GeTe. (A) Computed vibrational frequencies as a function of external pressure: the change is small for graphite, Si, and NiAl. (B) Computed mode-specific Grüneisen parameters for these materials, measuring the change of frequencies with pressure. A modest value of $\approx 2$ is obtained for NiAl; the results for the other two materials are even lower. (C, D) Same for GeTe. Here, a clearly different behavior is observed: note the large absolute values of the mode-specific Grüneisen parameters and the divergence around the pressure-induced rhombohedral → cubic transition. These data show that metavalent bonding, as found in GeTe, is not merely an intermediate between covalent and metallic bonding (or an extreme case of either), but should be viewed as an interaction mechanism in its own right.

We investigate this in **Fig. 2A**, where we used first-principles calculations to obtain the frequencies of vibrational modes in prototypical materials. We performed density-functional theory (DFT) calculations with the PBE functional (*20*) and norm-conserving pseudopotentials as implemented in ABINIT (*21*). Energy cutoffs and *k*-point grids were set to mirror the Materials Project standards (*22*), and phonons were obtained using density-functional perturbation theory including local field effects. Using DFT for this purpose has been validated against an inelastic X-ray scattering experiment on graphite before (*23*). We are here most interested in the *relative* change of the phonon frequencies with pressure, and Fig. 2A already shows that this change is not strongly pronounced. To quantify this, we calculate the mode-specific Grüneisen parameters $\gamma_i$. The latter is a dimensionless quantity that measures the volume dependence of the frequency $\omega_i$ for a given vibrational mode in the lattice, and thereby characterizes the anharmonicity of the interaction potential:

$$\gamma_i = -\frac{V}{\omega_i}\frac{\partial \omega_i}{\partial V}.$$

The overall magnitude of these mode-specific Grüneisen parameters in graphite is small, and slightly higher in a prototypical covalent solid, Si, and in the metal NiAl (**Fig. 2B**). In no case do we observe very pronounced changes under pressure.

By contrast, GeTe behaves very differently both in qualitative and quantitative terms (**Fig. 2C**). The absolute values of the Grüneisen parameters are much larger, indicating a higher sensitivity. By exerting pressure on GeTe, the structure approaches the ideal rocksalt type ($R3m \rightarrow Fm\bar{3}m$ transition), and therefore pressure serves as a direct means to control the extent of the Peierls distortion. At the same time, the system exhibits a band gap at ambient pressure, but has metallic occupation at high pressure,[1] and the structural transition is linked to an electronic instability (reflected in an anomalous increase in both Born effective charges and dielectric constants; Fig. S1). The large magnitude of the $\gamma_i$ (**Fig. 2D**) shows that the interaction potential for GeTe is very anharmonic. Furthermore it explains why materials like GeTe have such a low thermal conductivity (*14, 24*); since high values of the mode-specific Grüneisen parameter lead to low thermal conductivities of the lattice. As a consequence, incipient metals have been shown to be promising candidates for thermoelectrics (*14, 25*). No similar effect is observed in graphite or graphene.

Finally, the above-mentioned concept of electron deficiency in (incipient) metals does not apply to systems with conjugated π systems either—indeed, benzene and graphene have *more* electrons than those in their $sp^2$ backbone. Together with the arguments above, this underlines the fundamental difference between both material classes. Use of the term "resonant bonding" should therefore be restricted to benzene, graphene, and its analogues (**Fig. 3**).

[1] The data presented in Fig. 2 was obtained using metallic occupation of states for cubic GeTe, graphite, and NiAl, while semiconductor-like occupation was used for silicon and rhomboedral GeTe. The use of metallic occupation for cubic GeTe was compulsory as cubic (fcc) GeTe proved to be unstable against Peierls distortion when semiconductor-like occupation was imposed.

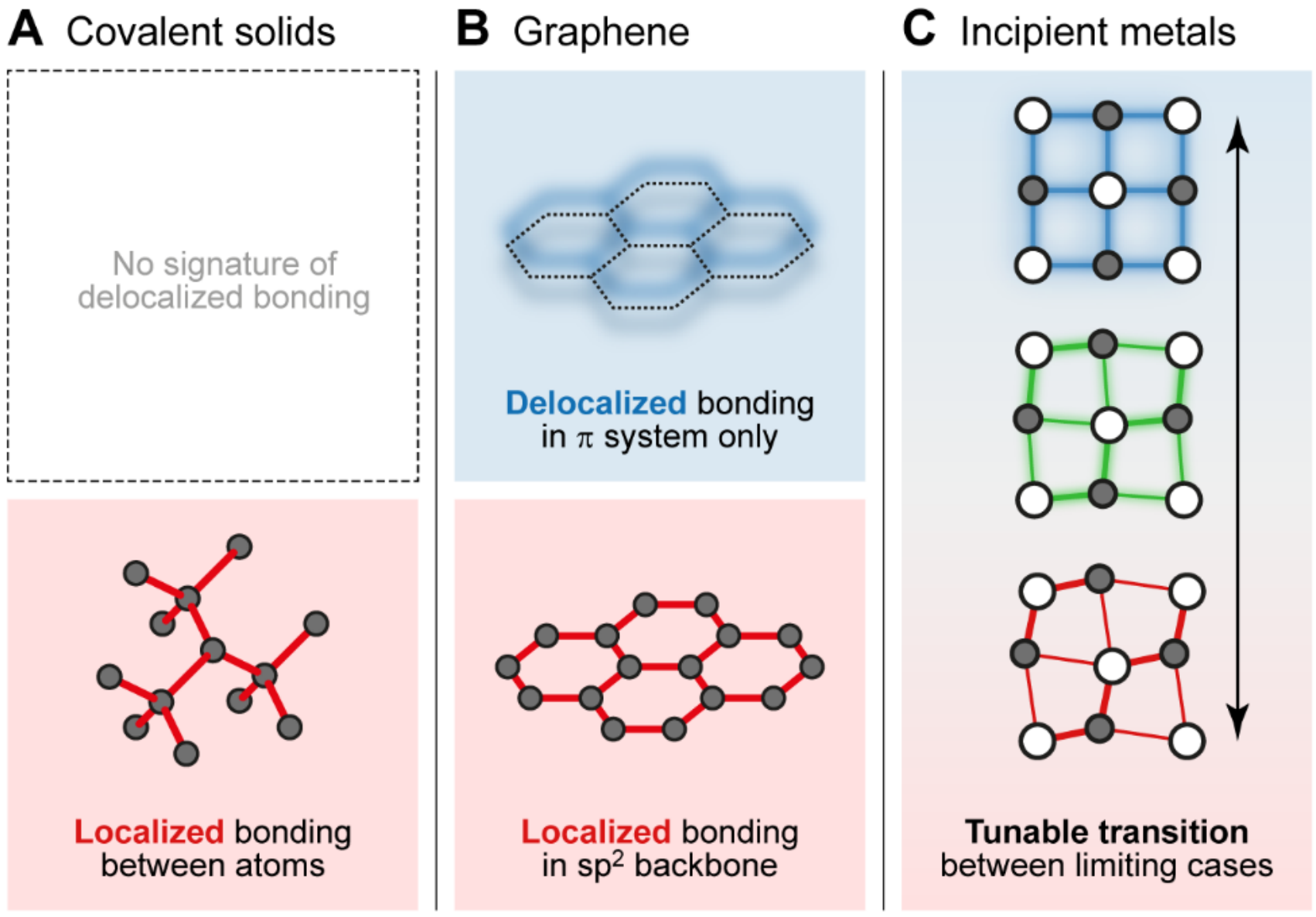

**Fig. 3**. Schematic overview of typical covalent and resonantly bonded systems, as well as incipient metals. (A) Covalent solids, say diamond-type silicon, form strong and localized bonds only. (B) In graphene and graphite, both localized $sp^2$ bonding (red) and delocalized $\pi$ bonding (blue) exist—but the corresponding orbitals are orthogonal, so the two bonding mechanisms exist side by side and largely independent from one another. (C) By stark contrast, incipient metals show a gradual transition, due to their unique bonding mechanism: they exhibit features both of localized and delocalized bonding, and can be tuned between both limiting cases (e.g., as shown in Fig. 2, through pressure-induced structural transformations).

We can now go one step further and derive a table that compares the three hitherto described fundamental bonding mechanisms in solids (metallic, covalent, and ionic) to our definition of metavalent bonding. To this end, we qualitatively assess all relevant properties or "fingerprints": the electric conductivity, the local atomic arrangement (characterized by the effective coordination numbers as above), the optical dielectric constant, the Born effective charge, and finally, the mode specific Grüneisen parameter for transverse optical modes. A comparison of these characteristic indicators reveals that metavalent bonding is a genuine mechanism of bonding in solids and not merely a combination (or an intermediate) of covalent and metallic bonding.

**Table 1.** Overview of characteristic materials fingerprints used to define bonding in solids. Numerical values for the corresponding quantities are collected in Table S1.

| | **Ionic** | **Covalent** | **Metavalent** | **Metallic** |
|---|---|---|---|---|
| Conductivity (**electrical fingerprint**) | Very low | Low | Moderate | High |
| Effective coordination number[a] (**structural fingerprint**) | 4 (ZnS), 6 (NaCl), 8 (CsCl) | 8–*N* rule typically satisfied | 8–*N* rule not satisfied | 8 (bcc), 12 (hcp / fcc) |
| Optical dielectric constant (**optical fingerprint**) | Low | Low | High | —[b] |
| Born effective charges (**chemical bond polarizability**) | Usually Low | Low | High | —[b] |
| Mode specific Grüneisen parameters (**anharmonicity**) | Usually Low | Low | High | Low |

[a]For ionic and metallic systems, representative (but not exclusive) example structure types are given.
[b]These indicators are not normally applicable to the metallic state.

These findings are not only of fundamental interest, but relevant for a very practical problem: they provide a new perspective on the functional principle of PCMs for data storage, which are switched between a crystalline phase (“one bits”) and an amorphous phase (“zero bits”) in applications (*26*). The coordination numbers in amorphous PCMs come much closer to obeying the (8–*N*) rule (*27*), albeit their complex structures will lead to a visible distribution of ECoNs in a given sample, and cases are known where chemical ordering effects modify this simple view of bonding (*28*). Indeed, amorphous PCMs show none of the above specific fingerprints of metavalent bonding (Table 1) (*12*, *29*), and are classically covalently bonded. Upon transition to the crystalline states, the bonding transcends classical covalency, which manifests itself in four ways: (i) the coordination numbers increase, such that the 8–*N* rule is no longer satisfied; (ii) the electronic polarizability rises sharply, which leads to a high optical dielectric constant and thus to the optical contrast between the amorphous and crystalline state; (iii) at the same time, the chemical bond polarizability rises, leading to unusually high Born effective charges (cf. Fig. 1B), and finally (iv) the vibrational properties show a pronounced change, including unusual phonon softening (*30*) and large values of the mode-specific Grüneisen parameters (Fig. 2). Hence, one can also characterize phase-change materials for data storage as "bond-change materials".

In conclusion, we have identified a unique bonding mechanism in a class of solid-state materials including PCMs: its characteristics are between those of covalency and metallicity, but distinctly different from both. We suggest replacing the currently used term "resonant bonding" by *metavalent bonding* for this class of materials, and to call them *incipient metals*. This avoids the previously ambiguous wording, as we have shown that "resonant bonding" in graphite and related π-conjugated systems is fundamentally different from metavalent bonding. We believe that incipient metals will provide an ideal playground for studying structure–bonding–property relationships in the future. This will be instrumental for the chemically guided discovery of new materials with unconventional property combinations, including PCMs and new candidates for thermoelectrics.

**Acknowledgments:** M.W. thanks S. Blügel for piquing his interest in the nature of "resonant bonding". We are grateful to R. Dronskowski, A. Tchougréeff, F. Ducastelle, R. Lobo, and C. Gatti for ongoing discussions. We thank F. Lange, M. Cagnoni, O. Cojocaru-Mirédin, M. Oliveira, and S. Jakobs for help with Fig. 1 and remarks on the manuscript, and S. Jakobs for bringing ECoNs to our attention and tabulating the corresponding data. M.W. acknowledges support by the DFG (SFB 917). Moreover, the research leading to these results has received funding from the European Union Seventh Framework Programme (FP7/2007-2013) under grant agreement no. 340698, as well as the Excellence Initiative (Distinguished Professorship). C.B. and J.-Y. R. acknowledge support from the French Research Funding Agency (ANR-15-CE24-0021-05, SESAME). J.-Y.R. acknowledges computational resources provided by the CÉCI funded by the F.R.S.-FNRS under Grant No. 2.5020.11 and the Tier-1 supercomputer of the Fédération Wallonie-Bruxelles, infrastructure funded by the Walloon Region under grant agreement n°1117545.

**Supplementary Materials:**

Figure S1
Table S1
Supplementary References S1–S20

**Supplementary Materials:**

**Pressure-dependent bonding indicators for GeTe.** Below we show computed values of Born effective charges (Z*; blue) and dielectric constants ($\varepsilon_\infty$; orange) for GeTe. Similar to the Grüneisen parameters (Fig. 2D in the main text), these evidence anomalous behavior and an electronic instability in the material that is not observed in any other of the prototype materials investigated.

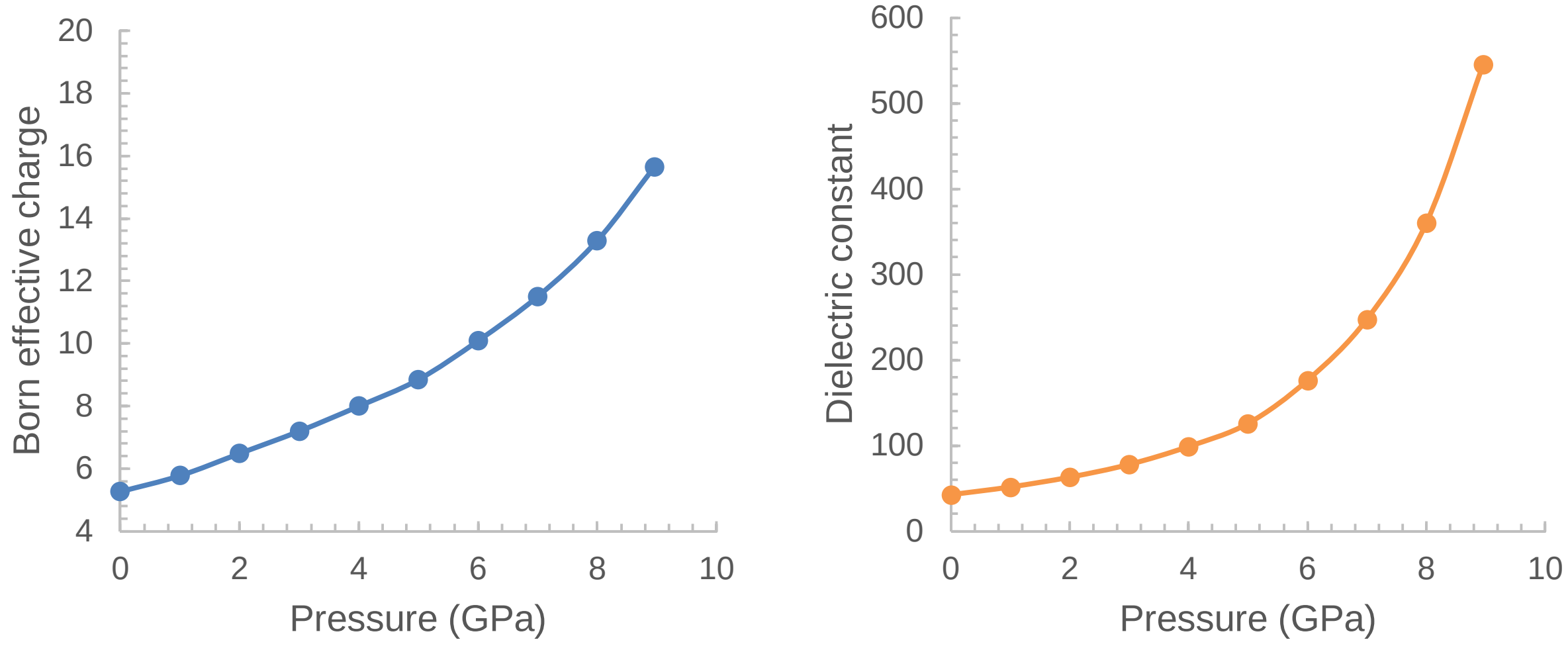

**Fig. S1**. Pressure dependence of relevant bonding indicators in rhombohedral GeTe, obtained from DFT computations as described in the main text. Lines are only guides to the eye.

**Tabulated data.** In the following, we detail the data used to generate the plots in Fig. 1 in the main text. Conductivity values have mainly been extracted from the Springer Materials database and the underlying original references, listed as Supplementary References (*S1*–*S20*) below. We do emphasize an inherent degree of uncertainty in this procedure, as data are necessarily collected from different experimental reports, and may depend on growth conditions, doping, etc.—wherever possible, we have chosen data for high-quality crystals without external doping. We also emphasize that, even with these differences, the physical trends observed in Figure 1 in the main text remain unchanged. Born effective charges *Z** were computed using density-functional perturbation theory as detailed in the manuscript; here, all computations are done in the same framework, so that the data are directly comparable. Effective coordination numbers ECoN have been obtained following Brunner and Schwarzenbach (Ref. (*19*) in the manuscript).

**Table S1.** Physical properties of materials included in Fig. 1 in the main text; in addition, more examples of mode-specific Grüneisen parameters $\gamma$ for transverse optical modes are provided.

| Species | Bonding | conductivity σ (S / cm) | Ref. | log σ | *Z** | ECoN | $\gamma_{TO}$ |
|---|---|---|---|---|---|---|---|
| $AgSbTe_2$ | *metavalent* | $1.6 \times 10^{2}$ | S1 | 2.20 | 4.30 | 6 | 3.03 |
| PbSe | *metavalent* | $2.4 \times 10^{2}$ | S2 | 2.38 | 4.81 | 6 | |
| $Sb_2Te_3$ | *metavalent* | $2.3 \times 10^{3}$ | S3 | 3.36 | 4.61 | 5.76 | |
| $Bi_2Se_3$ | *metavalent* | $1.0 \times 10^{3}$ | S2 | 3.00 | 3.97 | 5.76 | |
| $Bi_2Te_3$ | *metavalent* | $6.6 \times 10^{2}$ | S4 | 2.82 | 4.89 | 5.76 | |
| GeTe | *metavalent* | $5.0 \times 10^{3}$ | S3 | 3.70 | 6.06 | 5.31 | |
| SnTe | *metavalent* | $9.8 \times 10^{3}$ | S5 | 3.99 | 6.07 | 6 | 5.83 |
| PbTe | *metavalent* | $2.9 \times 10^{3}$ | S6 | 3.47 | 5.70 | 6 | 15.6 |
| Hg | metallic | $4.8 \times 10^{4}$ | S7 | 4.69 | — | 7.3 | |
| Po | metallic | $2.4 \times 10^{4}$ | S7 | 4.38 | — | 6 | |
| Pb | metallic | $5.2 \times 10^{4}$ | S7 | 4.71 | — | 12 | |
| Li | metallic | $1.2 \times 10^{5}$ | S7 | 5.07 | — | 8 | |
| Ni | metallic | $1.6 \times 10^{5}$ | S7 | 5.21 | — | 12 | |
| Zn | metallic | $1.8 \times 10^{5}$ | S7 | 5.26 | — | 12 | |
| W | metallic | $2.1 \times 10^{5}$ | S7 | 5.32 | — | 8 | |
| Al | metallic | $4.1 \times 10^{5}$ | S7 | 5.61 | — | 12 | |
| Au | metallic | $4.9 \times 10^{5}$ | S7 | 5.69 | — | 12 | |
| Cu | metallic | $6.5 \times 10^{5}$ | S7 | 5.81 | — | 12 | |
| Ag | metallic | $6.8 \times 10^{5}$ | S7 | 5.83 | — | 12 | |
| $Sb_2Se_3$ | p-bonded | $4.0 \times 10^{-7}$ | S8 | –6.40 | 3.60 | 3.92 | |
| GeSe | p-bonded | $1.3 \times 10^{-6}$ | S9 | –5.89 | 2.96 | 3.09 | 1.41 |
| SnSe | p-bonded | $2.5 \times 10^{-5}$ | S9 | –4.60 | 3.49 | 3.13 | 1.97 |
| $Sb_2S_3$ | p-bonded | $1.0 \times 10^{-8}$ | S10 | –8.00 | 3.39 | 3.92 | |
| SnS | p-bonded | $3.4 \times 10^{-4}$ | S11 | –3.46 | 3.44 | 3.43 | |
| ZnS | $sp^3$ | $1.0 \times 10^{-6}$ | S12 | –6.00 | 2.00 | 4 | 1.80 |
| GaAs | $sp^3$ | $1.0 \times 10^{-8}$ | S13 | –8.00 | 2.20 | 4 | 1.21 |
| Si | $sp^3$ | $1.5 \times 10^{-8}$ | S14 | –7.81 | 0.00 | 4 | |
| Ge | $sp^3$ | $3.3 \times 10^{-2}$ | S15 | –1.48 | 0.00 | 4 | 1.14 |
| InSb | $sp^3$ | $2.2 \times 10^{2}$ | S16 | 2.34 | 2.50 | 4 | 1.41 |
| CdS | $sp^3$ | $5.9 \times 10^{-6}$ | S17 | –5.23 | 2.21 | 4 | |
| AlSb | $sp^3$ | $1.2 \times 10^{0}$ | S16 | 0.08 | 1.89 | 4 | |
| InAs | $sp^3$ | $5.0 \times 10^{1}$ | S16 | 1.70 | 2.74 | 4 | 1.33 |
| InP | $sp^3$ | $5.0 \times 10^{0}$ | S16 | 0.70 | 2.74 | 4 | |
| AlAs | $sp^3$ | $9.5 \times 10^{0}$ | S16 | 0.98 | 2.15 | 4 | |
| GaSb | $sp^3$ | $\approx 10^{3}$ | S18 | 3.00 | 1.91 | 4 | 1.23 |
| HgSe | $sp^3$ | $6.0 \times 10^{3}$ | S19 | 3.77 | 3.38 | 4 | |
| HgTe | $sp^3$ | $9.3 \times 10^{2}$ | S20 | 2.97 | 3.30 | 4 | |

## Supplementary References